\documentclass[12pt]{iopart}
\usepackage{iopams}
\usepackage{graphicx}

\begin{document}

\title{Controlled MOCVD growth of Bi$_2$Se$_3$ topological insulator nanoribbons}

\author{L. D. Alegria}
\author{J. R. Petta}
\address{Department of Physics, Princeton University, Princeton, NJ 08544, USA}
\address{Princeton Institute for the Science and Technology of Materials, Princeton University, Princeton, NJ 08544, USA}
\ead{petta@princeton.edu}

\begin{abstract}
Topological insulators are a new class of materials that support topologically protected electronic surface states. Potential applications of the surface states in low dissipation electronic devices have motivated efforts to create nanoscale samples with large surface-to-volume ratios and highly controlled stoichiometry. Se vacancies in Bi$_2$Se$_3$ give rise to bulk conduction, which masks the transport properties of the surface states. We have therefore developed a new route for the synthesis of topological insulator nanostructures using metalorganic chemical vapour deposition (MOCVD). MOCVD allows for control of the Se/Bi flux ratio during growth. With the aim of rational growth, we vary the Se/Bi flux ratio, growth time, and substrate temperature, and observe morphological changes which indicate a growth regime in which nanoribbon formation is limited by the Bi precursor mass-flow. MOCVD growth of Bi$_2$Se$_3$ nanostructures occurs via a distinct growth mechanism that is nucleated by gold nanoparticles at the base of the nanowire. By tuning the reaction conditions, we obtain either single-crystalline ribbons up to $10$ $\mu \rm m$ long or thin micron-sized platelets.
\end{abstract}


\submitto{\NT}

\maketitle

\section{Introduction}

Topological insulators (TIs) are bulk insulators possessing helical surface states that span the bulk band gap as a consequence of strong spin-orbit coupling \cite{Hasan2010,Qi2010a, Kane2005, Fu2007}. While Bi$_2$Se$_3$ shares these properties with a set of Bi and Sb chalcogenides, it is unique among the 3D topological insulators as it has a comparatively large band gap of 0.35 eV and a simple surface spectrum consisting of a single Dirac cone roughly centered within the gap \cite{Hasan2010, Greenaway1965}. These features have made it a system of choice for TI experiments \cite{Roushan2009, Xia2009}.

To date, many of the most informative experiments have been performed using surface sensitive probes, such as angle resolved photoemission spectroscopy (ARPES) or scanning tunnelling microscopy (STM) \cite{Roushan2009, Xia2009}. Potential applications of TI compounds will require the development of transport measurements, which so far have been hindered by bulk conduction in these materials, which masks the transport properties of the surface states \cite{Checkelsky2009}. In Bi$_2$Se$_3$, Se-vacancies typically result in n-type doping that pins the Fermi level in the bulk conduction band \cite{Greenaway1965, Hor2009}. Several approaches are being explored to tune the Fermi level into the gap. In particular, nanoscale samples offer an increased surface-to-volume ratio and have allowed the employment of high-k dielectrics or electrolytic solutions for field effect gating \cite{Checkelsky2011, Peng2010,Xiu2011, Kim2012, Steinberg2011}. However, very high electric fields can induce dielectric breakdown, band bending, and in the case of polymeric gating, undesired chemical reactions at the surface of the crystal \cite{Kim2012, Efetov2010}. Therefore it is ideal to eliminate Se-vacancies to the greatest extent possible. While nanoscale samples may be exfoliated from very high purity macroscopic crystals, there is typically some degradation relative to the bulk host material \cite{Checkelsky2011}. Together, these factors encourage the development of direct syntheses of nanoscale Bi$_2$Se$_3$ with minimal defects.

In numerous other materials, the vapour-liquid-solid (VLS) growth mechanism has afforded the possibility of epitaxial growth of high quality structures with predefined geometries \cite{Fan2006, Rao2003, Kolasinski2006}. VLS growth of Bi$_2$Se$_3$ nanostructures has previously been demonstrated using a convenient solid-source growth method by Kong \textit{et al.} \cite{Kong2010a}. In this approach, an inert gas flows over a heated Bi$_2$Se$_3$ powder, carrying Bi and Se vapours to a cooler growth substrate downstream.  Liquid Au particles on the substrate saturate with the vapour and precipitate solid nanowires and nanoribbons. Au particles remaining at the tips of the ribbons are evidence for the VLS growth process. In such a configuration, however, it is difficult to control the magnitude and time dependence of the Bi and Se vapour concentration \cite{Cafaro1984}. Despite recent work utilizing nanoribbons for TI transport experiments, there has been comparatively little work studying the growth process itself \cite{Peng2010,Kong2010a,Hong2012}. In conventional semiconductors, such as Si, the VLS growth process has been studied in-situ using chemical vapour deposition (CVD) in concert with high resolution TEM imaging \cite{Kolasinski2006,Ross2010}. Given the degree of control of the doping level that is required for successful TI transport measurements, this class of compounds would benefit from detailed studies of the growth mechanism.

In this work we employ a metalorganic chemical vapour deposition (MOCVD) growth process in which Bi$_2$Se$_3$ nanostructures are grown in a controlled environment with well-defined, independently adjustable Bi and Se vapour concentrations.  In our study we first obtain nanostructure growth at a specific set of growth parameters and then vary the parameters in order to understand the process of nanoribbon formation. The ratio of Bi to Se vapour is of particular importance in the interest of reducing Se vacancies and is the focus of our study.

\section{Methods}

We perform growth in a MOCVD system depicted schematically in figure 1. The reactor is based on a design that has yielded high quality growth of InP and InAs nanowires \cite{Schroer2010a}. The growth chamber consists of a 12 inch spherical chamber that has a base pressure $<5\times10^{-10}$ Torr. Samples are loaded into the chamber via a sample load lock that is evacuated to $1 \times 10^{-6}$ Torr.  The substrate rests on a 2 inch boron nitride heater stage at the chamber center and the temperature is monitored using a standard thermocouple. Mass flow controllers (MFCs) admit prescribed flow rates of hydrogen through bubblers containing the liquid metalorganic precursors diethyl selenium (DESe) and trimethyl bismuth (TMBi). The metalorganic vapours enter a cold-wall growth chamber where they decompose at the heated substrate. DESe and TMBi are commercially available in high purities, have cracking temperatures of $\sim$ 450 $^\circ$C, and are unlikely to react in the gas phase \cite{Giani2002, Seshan2002}. Among possible Se precursors, DESe has several advantages, including atmospheric stability and low toxicity.

\begin{figure}
\includegraphics[width=6.55 in]{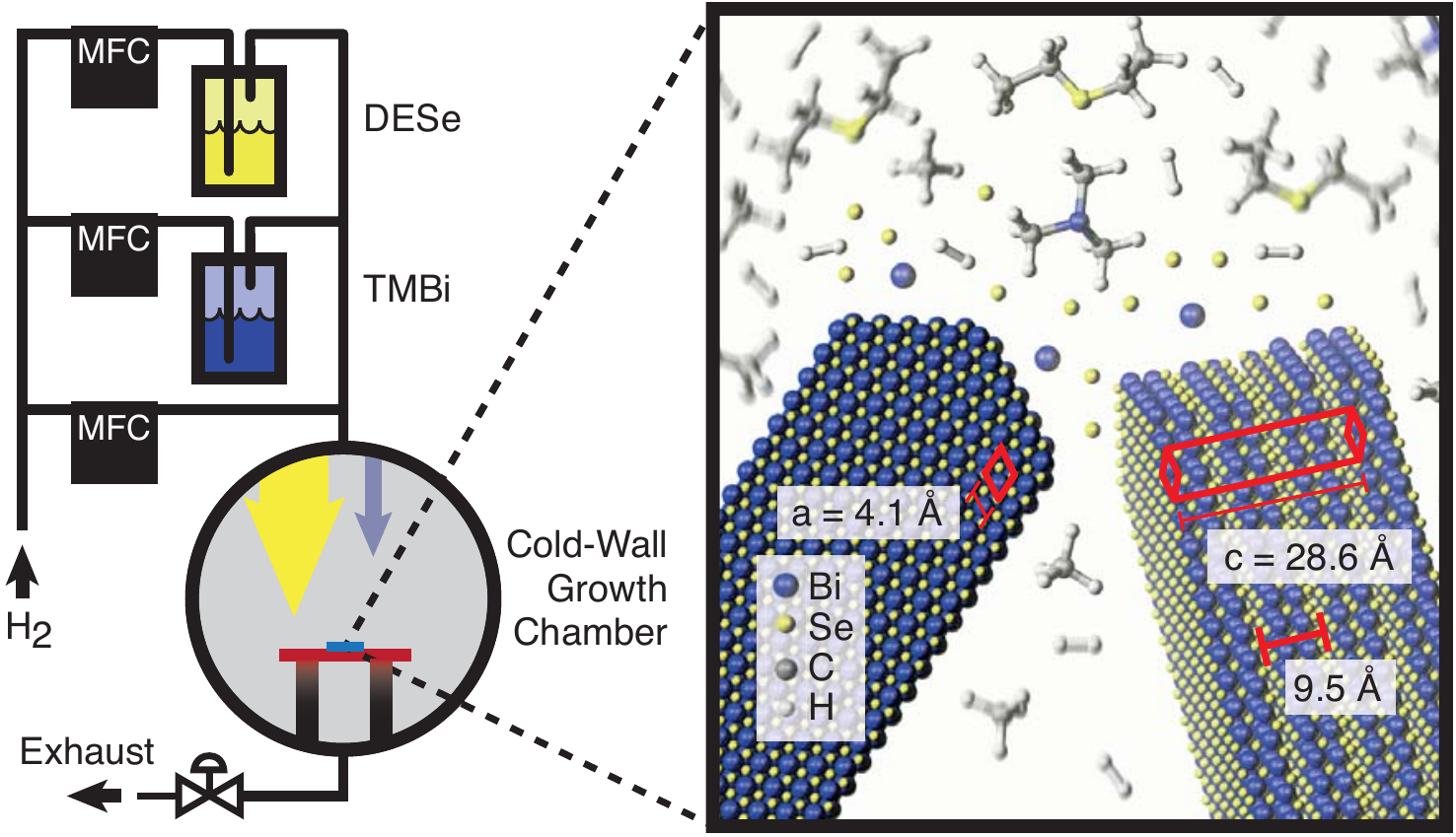}
\caption{\label{fig1} Schematic illustration of the growth process. Hydrogen carrier gas passes through bubblers and delivers a dilute vapour of trimethyl bismuth and diethyl selenium to the heated substrate (\emph{left}).  At the substrate the precursors decompose, forming long crystallites of  Bi$_2$Se$_3$ (\emph{right}).  Bi$_2$Se$_3$ belongs to the $D^5_{3d}$ space group and the unit cell (a = b = 4.135 \AA, c = 28.615 \AA) is highlighted in red. The structure consists of stacked quintuple layers with 9.5 \AA \ thickness. Viewed along the c-axis and nearly along the a-b plane, the hexagonal, layered structure is apparent.  Atoms within a quintuple layer are covalently bonded, whereas neighbouring quintuple layers are Van der Waals bonded.}
\end{figure}

To obtain reproducible growth, the temperatures of the TMBi and DESe bubblers are maintained at 6.0 $^\circ$C and 25.0 $^\circ$C, respectively. The well-established vapour pressures of the liquids and the precise H$_2$ flows through the MFCs allow for straightforward calculation of the partial pressures in the growth chamber \cite{Seshan2002}.  A TMBi partial pressure $p_{\rm{TMBi}} = 1 \times 10^{-5}$ atm is used and the DESe/TMBi partial pressure ratio, $r$, is varied in the range $r$ = 3--45.  An excess of Se flux is necessary to compensate for the Se evaporating from the crystal during growth \cite{Seshan2002}.  Precursor vapour pressures and constraints on H$_2$ flows prevent us from exceeding $r$ = 45 while maintaining fixed TMBi pressure.

Prior studies of Bi$_2$Se$_3$ MOCVD thin film deposition using DESe and TMBi serve as a guide for the above growth conditions, although differences are expected for nanoribbon growth \cite{Giani2002, AlBayaz2002, AlBayaz2003}.  These studies report the onset of mass-flow limited deposition above about 465 $^\circ$C, although at a significantly higher flow rate than we use here.  In the present study, the growth substrate temperature is typically maintained at 470 $^\circ$C to ensure complete decomposition of the precursors.  In addition, thin film studies observed Se deficient films for ratios below about 20, which led us to typically use a ratio near 30.

Bi$_2$Se$_3$ nanostructures are grown on (100) orientation Si substrates that are prepared with a 5 nm thick thermally evaporated Au film.  At elevated temperatures the film forms a dense layer of Au nanoparticles 10--40 nm in diameter (see figure 8). Several other substrates including SiO$_2$, Si (111), and GaAs were also tested, but produced no ribbon growth. In an alternative preparation, substrates are immersed in poly-l-lysene, blown dry and immersed in colloids of 60 nm Au nanoparticles, producing a layer of Au particles spaced $\sim$1 $\mu$m apart on the surface.

MOCVD allows for precise control of the entire growth process.  Once an overall H$_2$ carrier gas flow is initiated (600 sccm), active pressure control stabilizes the growth pressure to 100 Torr.  The sample is heated to the growth temperature (460--500 $^\circ$C) in 30 seconds. Precursor flow is initiated when the sample heater stabilizes to within 0.1 $^\circ$C of the prescribed growth temperature. Precursor flows are maintained for $t$ = 7--30 minutes for sample growth. Flow of the Bi precursor is halted to terminate growth. The sample heater is then turned off and the sample cools to 150 $^\circ$C in $\sim$200 seconds. Se evaporation from the nanoribbons is minimized by maintaining the flow of DESe until the sample has cooled to a temperature less than 150 $^\circ$C.

In the present study we systematically explore Bi$_2$Se$_3$ nanostructure growth under a range of conditions. We vary the growth substrate, growth duration, substrate temperature, and precursor ratio. As-grown samples are imaged using scanning electron microscopy (SEM). Sample thicknesses are determined by transferring growth products to Si substrates for subsequent imaging using an atomic force microscope (AFM). Crystal structure and chemical composition are determined using high resolution transmission electron microscopy (HRTEM) and energy dispersive spectroscopy (EDS).

\section{Results and Discussion}

A uniform layer of suspended nanoribbons is observed under a variety of growth conditions.  Bi$_2$Se$_3$ preferentially forms such thin, hexagonally faceted structures as a consequence of its crystal structure, which belongs to the space group $D^5_{3d}$ ($R3\bar{m}$) and consists of repeated quintuple layers (QLs), each containing five atomic layers in the order Se-Bi-Se-Bi-Se \cite{PerezVicente1999}. Weak van der Waals bonding between layers causes Bi$_2$Se$_3$ to grow along the basal plane defined by these layers, leading to the thin structures observed.  Clear trends in the growth morphology of these structures are observed as individual reaction conditions are varied.

Growth is significantly promoted on Si (100) substrates prepared with an evaporated 5 nm thick Au film.  Notably, ribbon growth is absent on Si (100) substrates lacking the Au layer under the above growth conditions.  We observe a lower density of growth when using colloidal Au nanoparticles, and growth is not apparent on other types of substrates under the same growth conditions.  Ribbons consistently grow `on-edge' (i.e. normal to the substrate and along the basal plane of the Bi$_2$Se$_3$ crystal) suggesting that nucleation depends on lattice-matching between Bi$_2$Se$_3$ and substrate.

High-yield growth occurs for a growth time $t$ = 15 minutes, substrate temperature $T$ = 470 $^\circ$C, and DESe/TMBi partial pressure ratio $r$ = 33.  From these values and the conditions described in the methods section, we independently vary $t$ = 7--30 minutes, $T$ = 460--500 $^\circ$C, and $r$ = 3--45.  In all these runs, we use centimetre sized Si (100) chips prepared with 5 nm thick Au films.

We first determine how the growth time affects nanowire growth. Figure 2 shows SEM images of nanowires grown at a reaction pressure $P$ = 100 Torr, with $T$ = 470 $^\circ$C, and $r$ = 30. There is a substantial increase in the average size of the nanoribbons upon increasing the growth time from 7 to 15 minutes. A further increase in growth time, up to 30 minutes, yields longer nanowires, but eventually a significant fraction of them fall over.

\begin{figure}
\includegraphics[width=6 in]{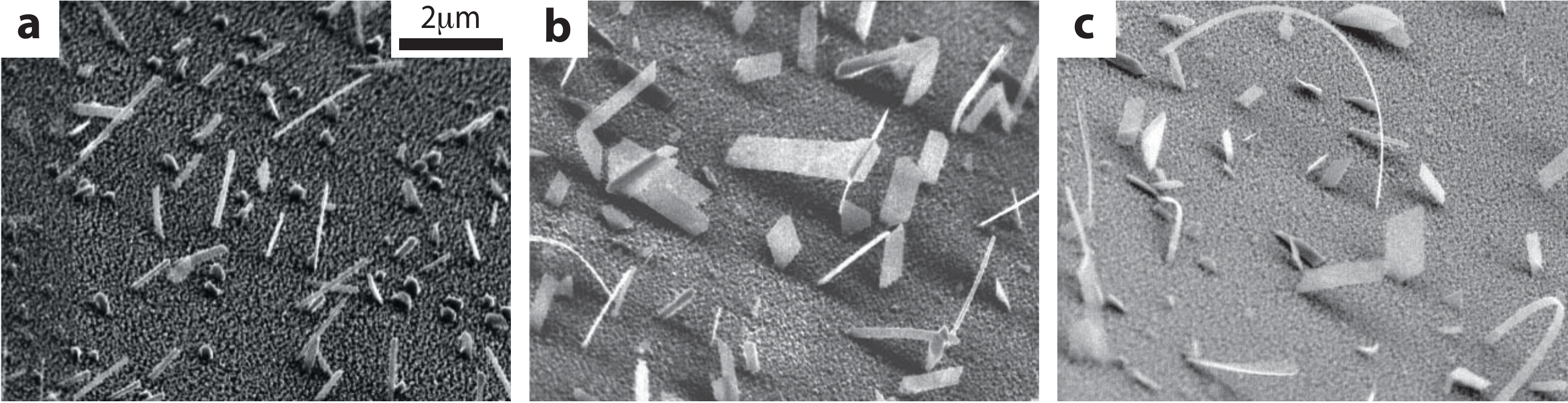}
\caption{\label{fig2} SEM images of growth products for growth durations of (a) 7.5 minutes, (b) 15 minutes, and (c) 30 minutes.}
\end{figure}

Figure 3 shows SEM images of nanostructures grown as a function of increasing growth temperature, which has a large impact on the structure of the nanowires. Very little growth is observed below 460 $^\circ$C. As the temperature is increased, growth product appears at the sites of the Au catalyst particles. A sudden onset of ribbon growth is observed at 470 $^\circ$C, consistent with a thermally activated decomposition process and previous reports in the literature indicating that the precursors fully decompose near this temperature \cite{AlBayaz2002}. The overall level of growth is roughly constant for temperatures above 480 $^\circ$C, suggesting a mass-flow-limited regime in which temperature does not appreciably change the growth product.

\begin{figure}
\includegraphics[width=6 in]{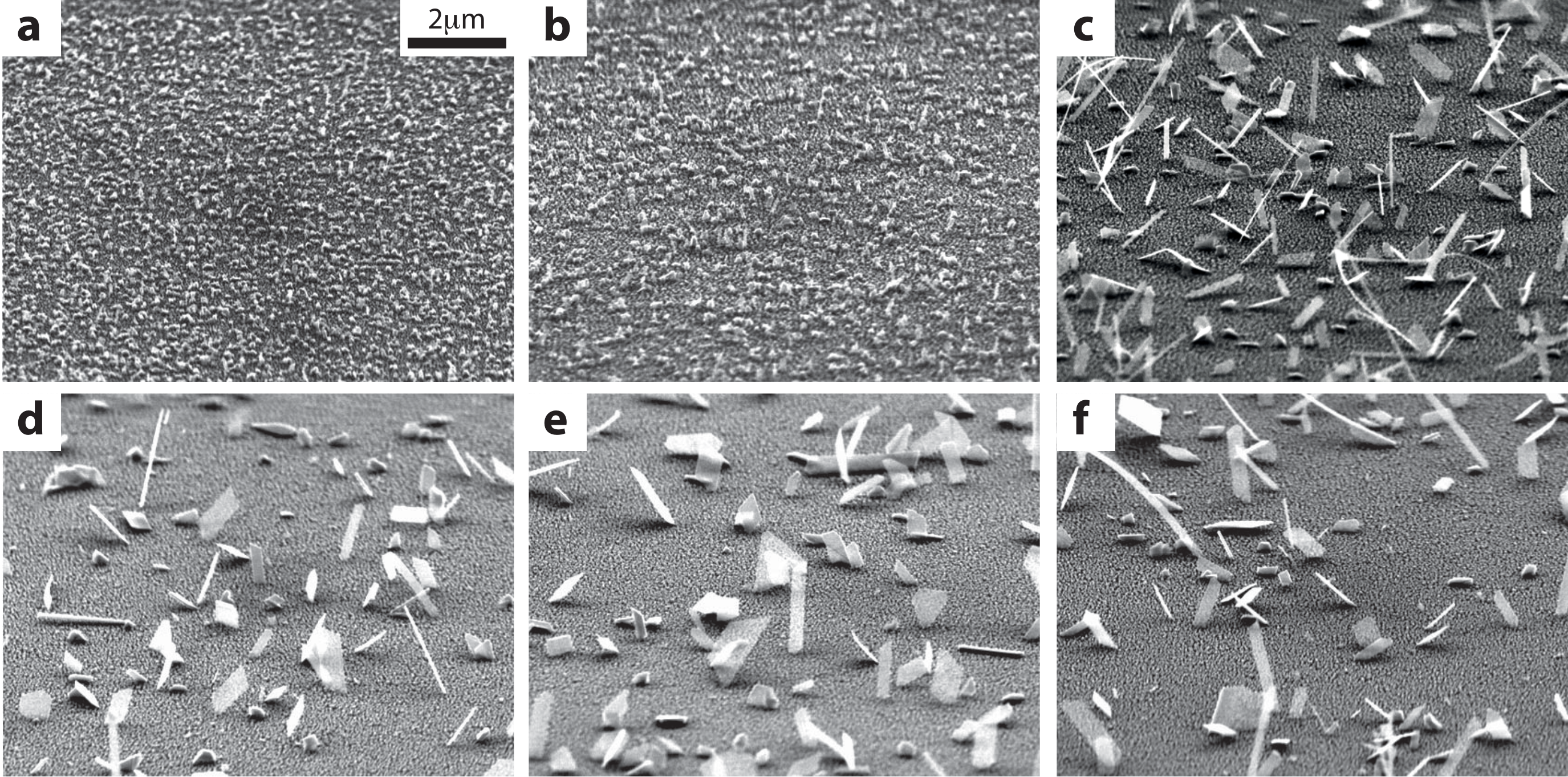}
\caption{\label{fig3} SEM images of growth products obtained at growth temperatures of (a) 460 $^\circ$C, (b) 465 $^\circ$C, (c) 470 $^\circ$C, (d) 480 $^\circ$C, (e) 490 $^\circ$C, and (f) 500 $^\circ$C.  Significant growth begins around 470 $^\circ$C, with little change in growth between 480-500 $^\circ$C, suggesting a mass-flow-limited regime of MOCVD.}
\end{figure}

Nanostructure yield is a sensitive function of the precursor partial pressure ratio $r$, as illustrated in figure 4. For this set of growth runs the precursor partial pressure ratio $r$ is varied holding a constant TMBi partial pressure = 1 $\times 10^{-5}$ atm. Reducing the ratio to 3 completely suppresses growth. At a ratio $r$ = 7, a small particulate appears, and at $r$ = 10 distinct nanoribbon growth begins. Overall growth increases between about 10--33 along with a gradual widening of ribbons.  Three growth runs up to $r$ = 45 displayed little or no change in product morphology.   The saturation above $r$ = 33 suggests a regime in which Se no longer limits growth.

\begin{figure}
\includegraphics[width=6 in]{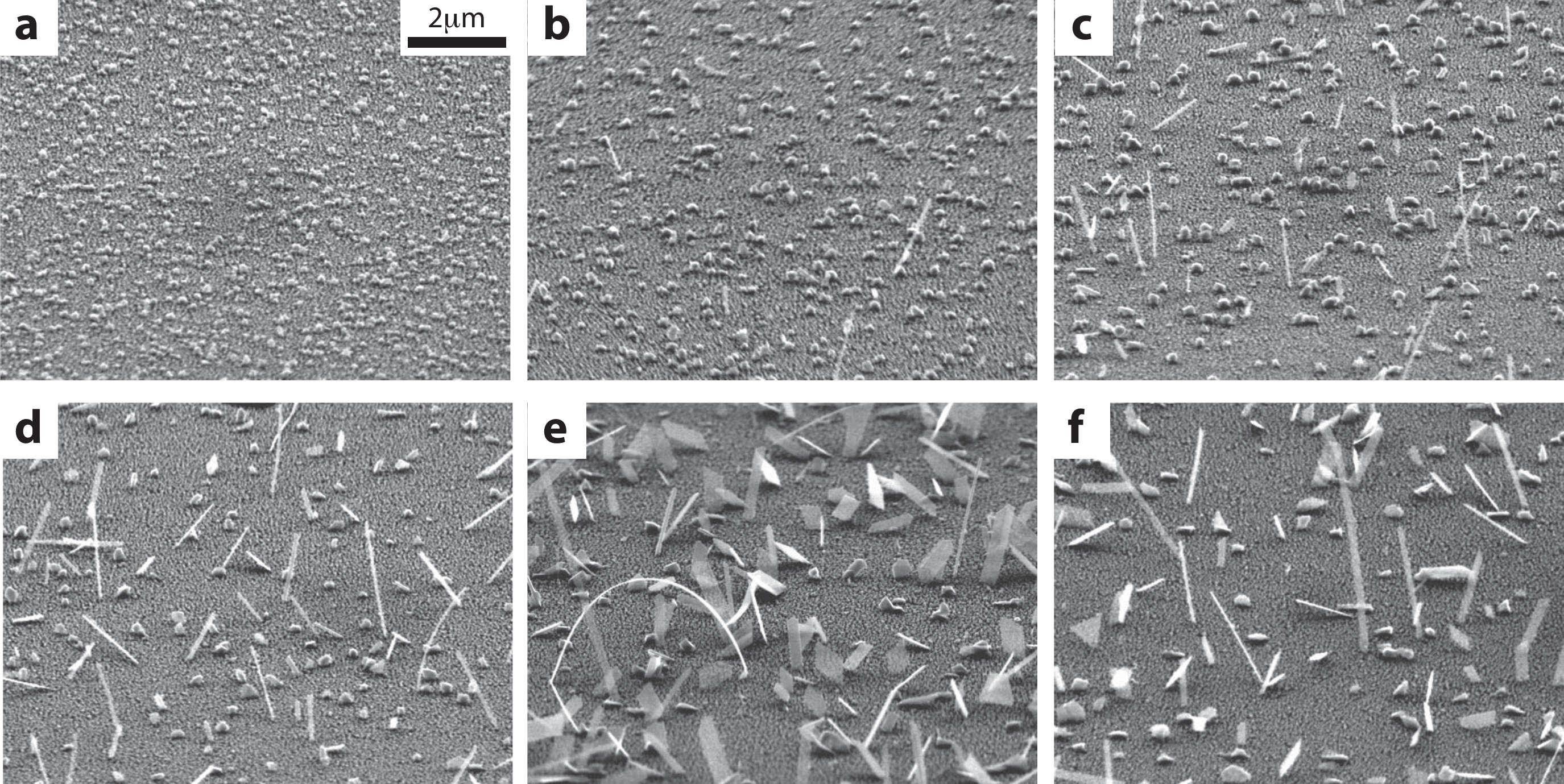}
\caption{\label{fig4} SEM images of growth products for precursor ratios (a) $r$ = 7, (b) 10, (c) 11, (d) 12, (e) 30, and (f) 33. Additional growth runs were performed up to $r$ = 45 with little change in growth morphology.  At $r$ $\sim$ 12, ribbons form with well-defined widths approximately equal to the Au catalyst particle diameter.}
\end{figure}

Growth products are quantitatively analyzed in figure 5. The areal density of nanowires with lengths $>$ 500 nm is plotted as a function of precursor partial pressure ratio in figure 5(a) (blue curve). The areal density of nanowires with widths $>$ 100 nm is plotted over the same range of parameter space (red curve). A strong dependence on $r$ is observed for $r$ $<$ 20, consistent with the SEM images shown in figure 4. The elemental composition of the ribbons is determined using TEM-based energy dispersive x-ray scattering. We characterize a set of five nanoribbons grown at $r$ = 30, showing the ribbons to be stoichiometric Bi$_2$Se$_3$ (see figure 5(b)).

\begin{figure}
\includegraphics[width=2.8 in]{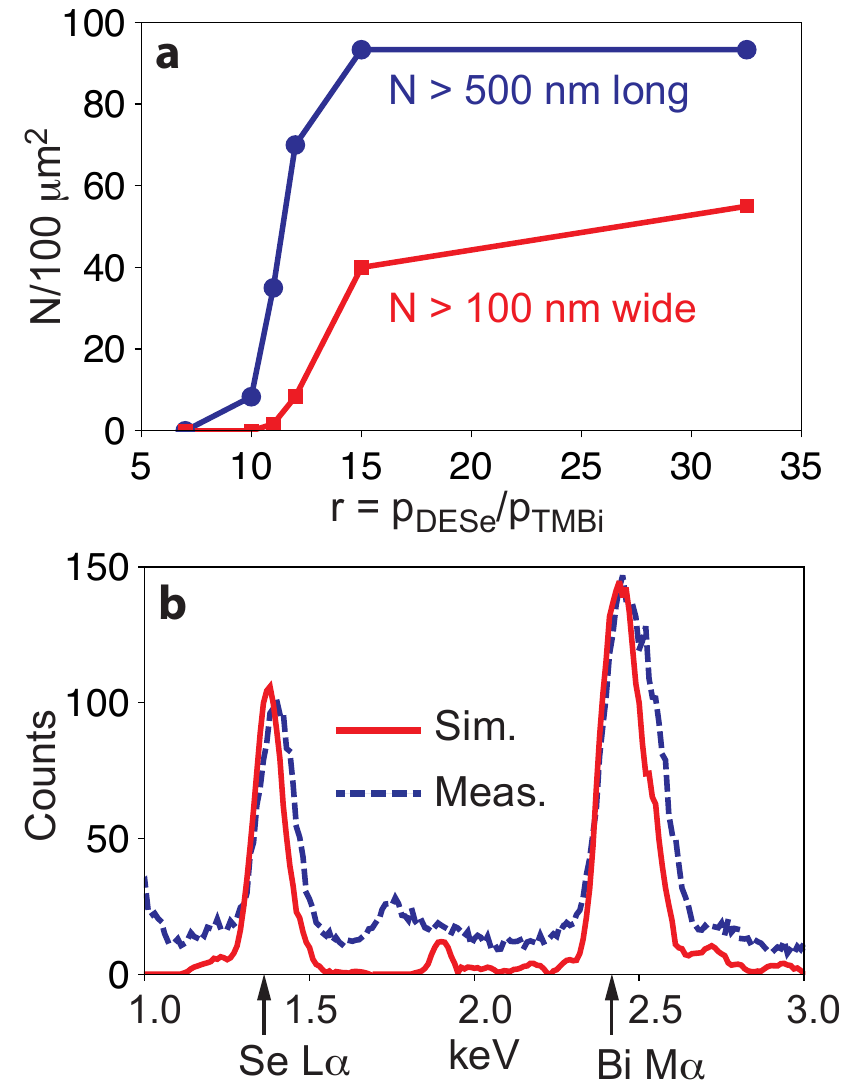}
\caption{\label{fig5} (a) Growth density as a function of precursor ratio, $r$. At fixed growth temperature, pressure, carrier gas flow, and duration (as specified in the text) wire growth begins when $r$ $\geq$ 7. Further increases in the precursor ratio result in wider and longer nanoribbons. Note that at $r$ = 33, longer wires are also present, with roughly 1/3 of the nanoribbons exceeding 1 $\mu$m in length. (b) Example of an EDS X-ray spectrum of a thin ribbon under a 200 keV electron beam.  The two peaks shown are used for measuring the Bi:Se ratio. The measured spectrum is shown by the dashed line and a Monte Carlo simulation (NIST DSTA-II) of a 10 nm film of Bi$_2$Se$_3$ is shown by the solid line.  Composition calculation is performed using the Evex software package.}
\end{figure}

Nanoribbons are also structurally characterized using SEM, AFM, and TEM.  The width and length are apparent in SEM images of as-grown chips, and occasional twisted ribbons show thicknesses $<$ 15 nm. Accurate determination of the nanoribbon thickness is best performed by AFM.  To do so, we briefly sonicate the growth chip in isopropanol, and place a droplet of the resulting suspension on an oxidized Si substrate, which is then blown dry with N$_2$. A study of 18 nanoribbons gave thicknesses of $20\pm5$ nm, some examples of which are shown in figure 6.

\begin{figure}
\includegraphics[width=2.8 in]{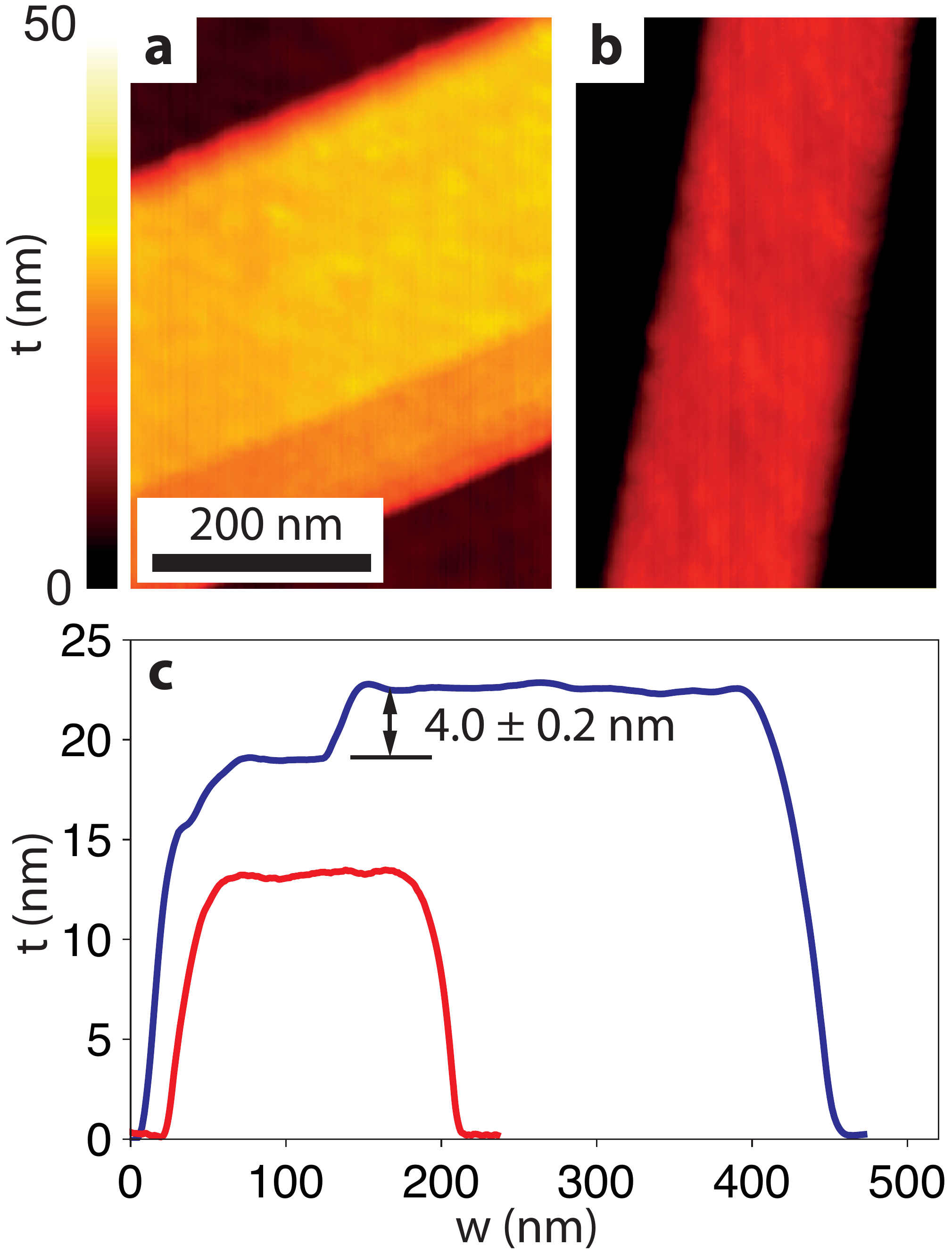}
\caption{\label{fig6} AFM measurements of the nanoribbons. (a) Nanoribbon with $\sim$ 23 nm thickness. There is a 4 QL step near the lower edge of the ribbon. (b) Nanoribbon with $\sim$ 13 nm thickness. (c) Thickness profiles averaged along the length of the wires in (a--b).}
\end{figure}

TEM provides high resolution structural information, as shown in figure 7. To image the ribbons we filter several droplets of the nanoribbon suspension through a porous carbon TEM grid. We determine crystal orientation by performing dark field imaging and selected area diffraction along the length of a nanoribbon. We observe growth exclusively along the (11$\bar{2}$0) direction. Using gold nanoparticles for calibration, we deduce a lattice spacing $a$ = 4.1 \AA  \hspace{0.1cm} from the diffraction pattern, in agreement with tabulated values in the literature \cite{PerezVicente1999}.  Clear lattice fringes are apparent in a 200 keV TEM, with a periodicity matching that of Bi$_2$Se$_3$.  The absence of fixed stress fringes within the crystallites indicates the ribbons are free of major defects. Many ribbons wrap around the edges of the pores in the grid, revealing thicknesses of $\sim$ 10 nm.   An amorphous region 1--5 nm wide along the edge of ribbons is visible, consistent with other's findings that a native bismuth and selenium oxide quickly forms on the surface of Bi$_2$Se$_3$ \cite{Kong2011}. 

\begin{figure}
\includegraphics[width=6 in]{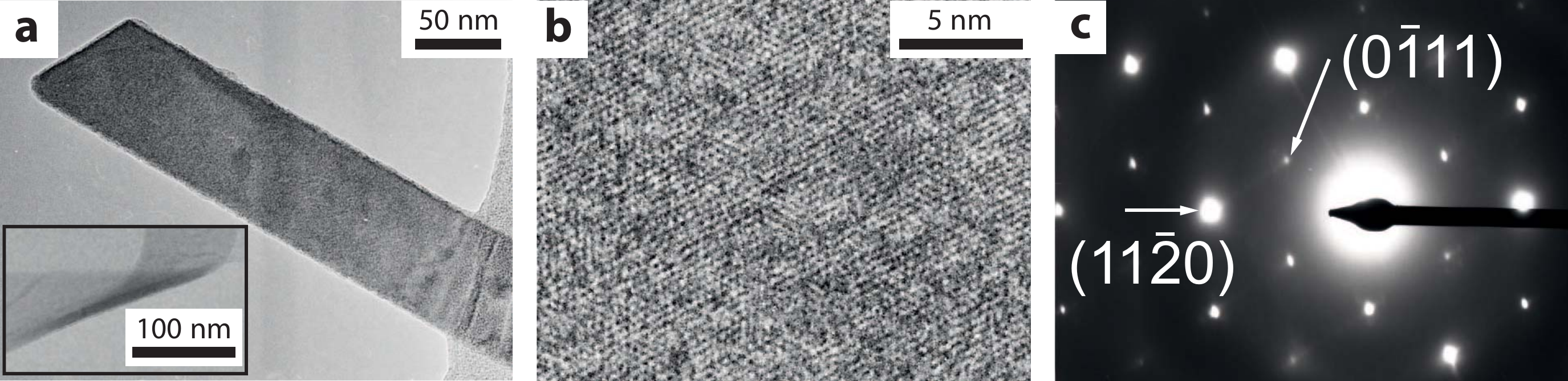}
\caption{\label{fig7} TEM studies confirm the single crystal nature of the wires, which are free of major crystalline defects.  (a) TEM image of the end of a ribbon suspended over a hole in the carbon grid.  Inset, bent wires, such as this one, allow accurate thickness measurements. (b)  HRTEM imaging shows a high degree of crystalline order. (c) Diffraction pattern from a flat region of a nanoribbon.}
\end{figure}

Images of as-grown substrates provide evidence for the growth mechanism.  Canonical VLS, as well as the solid-source Bi$_2$Se$_3$ nanowire growth as described in reference \cite{Kong2010a}, generally shows a Au particle at the free end of the nanowire, which is the particle from which the wire has precipitated.  In the samples described in the present work, a nanoparticle never appears at the top of the ribbon. Although it is possible that the liquid droplets have been exhausted by surface migration, it would seem likely that we would occasionally see a single particle that was not fully exhausted \cite{Hannon2006}.  On the other hand, it may be thought that a VS mechanism is taking place, in which nanoribbons grow directly from the vapour \cite{Kolasinski2006}.  However, this seems to be ruled out by the fact that ribbons do not grow on bare silicon, but require the presence of the Au.  A final possibility is that the Au catalyzes the decomposition of the metalorganic molecules into Bi and Se which then incorporate in the wires.  However, as noted, it is likely that precursors are thoroughly decomposed by the time they reach the substrate.

Although we cannot rule out second-order mechanisms, a natural conclusion is that a VLS mechanism is taking place, but growth is occurring from Au particles at the bottom of the ribbon.  A similar mechanism has been observed in other materials and is referred to as root-catalyzed growth \cite{Kolasinski2006}. The hypothesis is supported by the presence of Au particles at the bases of the nanoribbons, as illustrated in the SEM images shown in figure 8. It becomes clear that the catalyst particles are directly attached to the base of the nanoribbons once they are removed from the substrate, which allows for unambiguous SEM imaging. In addition, SEM images suggest that wider ribbons form by the participation of multiple nanoparticles at the crystallite vertices. We note that the substrate temperature used here is $\sim$100 $^\circ$C greater than the substrate temperature that is used in the solid-source method \cite{Kong2010a}. The elevated temperature is typically required for complete pyrolysis of DESe \cite{Stringfellow1989}.

\begin{figure}[t]
\includegraphics[width=3 in]{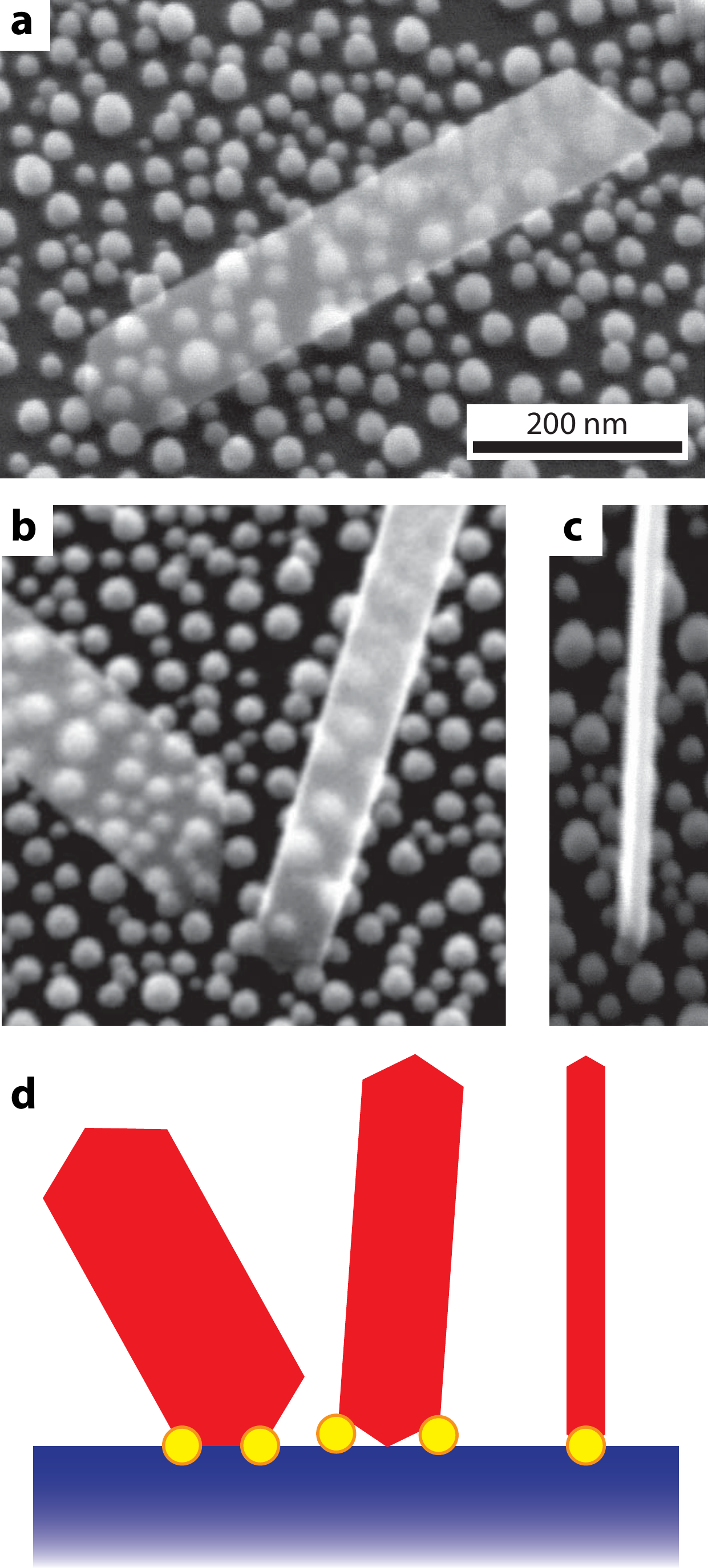}
\caption{\label{fig7} Growth mechanism: (a) Although Au is necessary for growth, a catalyst particle is never observed at the free end of the nanowire. (b) The presence of one or more Au particles at the base of each nanoribbon suggests that VLS growth occurs via a root-catalyzed process.  (c) High magnification images of the growth substrate consistently show Au particles at the vertices of large nanoribbons. Narrow nanoribbons appear to have widths that are set by the Au catalyst particle diameter. (d) Schematic of the root-catalyzed growth mechanism.}
\end{figure}

\section{Conclusions}
We demonstrate the synthesis of Bi$_2$Se$_3$ nanostructures using a tunable MOCVD growth process. Nanoribbons are grown in a VLS process using the precursors DESe and TMBi, with individual control of precursor partial pressures. Precise control of the reaction conditions afforded by MOCVD growth may allow the synthesis of Bi$_2$Se$_3$ nanoribbons with a reduced number of Se vacancies, enabling clearer transport measurements of the surface states found in TI compounds. Furthermore, we hope that the demonstrated control provides a pathway towards structurally deterministic synthesis of Bi$_2$Se$_3$ nanostructures, which may have significant implications for thermoelectric devices and TI-based spintronic devices.

\ack
We thank Anasua Chatterjee, Gerald Poirier, and Nan Yao at the NSF MRSEC funded Princeton Imaging and Analysis Center for assistance characterizing samples. Research was supported by the Sloan and Packard Foundations, and the NSF through the Princeton Center for Complex Materials, DMR-0819860.

\end{document}